\documentclass[12pt,a4paper,american]{article}
\usepackage[T1]{fontenc}
\usepackage[latin1]{inputenc}
\usepackage{palatino}
\usepackage{babel}
\usepackage{graphics}
\usepackage{setspace}
\makeatletter

\newcommand{\LyX}{L\kern-.1667em\lower.25em\hbox{Y}\kern-.125emX\@}
\newcommand{\noun}[1]{\textsc{#1}}
\let\SF@@footnote\footnote
\def\footnote{\ifx\protect\@typeset@protect
    \expandafter\SF@@footnote
  \else
    \expandafter\SF@gobble@opt
  \fi
}
\expandafter\def\csname SF@gobble@opt \endcsname{\@ifnextchar[
  \SF@gobble@twobracket
  \@gobble
}
\edef\SF@gobble@opt{\noexpand\protect
  \expandafter\noexpand\csname SF@gobble@opt \endcsname}
\def\SF@gobble@twobracket[#1]#2{}

\usepackage{cite}
\usepackage{amssymb}
\renewcommand{\vec}{\mathbf}

\renewenvironment{thebibliography}[1]
     {\section*{\refname
        \@mkboth{\MakeUppercase\refname}{\MakeUppercase\refname}}%
      \list{\@biblabel{\@arabic\c@enumiv}}%
           {\settowidth\labelwidth{\@biblabel{99}}%
            \leftmargin\labelwidth
            \advance\leftmargin\labelsep
            \@openbib@code
            \usecounter{enumiv}%
            \let\p@enumiv\@empty
            \renewcommand\theenumiv{\@arabic\c@enumiv}}%
      \sloppy
      \clubpenalty4000
      \@clubpenalty \clubpenalty
      \widowpenalty4000%
      \sfcode`\.\@m}
     {\def\@noitemerr
       {\@latex@warning{Empty `thebibliography' environment}}%
      \endlist}
\makeatother

\begin{document}

\def\a{1}

\title{\ifodd\a\else\vspace{-1cm}\fi Influence of magnetic fields on the spin
reorientation transition in ultra-thin films }

\author{A.~Hucht and K.~D.~Usadel\\
\textit{\small Theoretische Tieftemperaturphysik, Gerhard-Mercator-Universität,}\\
\textit{\small 47048 Duisburg, Germany}\small }

\date{~\vspace*{-1cm}}

\maketitle
\begin{abstract}
The dependence of the spin reorientation transition in ultra-thin ferromagnetic
films on external magnetic fields is studied. For different orientations
of the applied field with respect to the film, phase diagrams are calculated
within a mean field theory for the classical Heisenberg model. In particular
we find that the spin reorientation transition present in this model is
not suppressed completely by an applied field, as the magnetization component
perpendicular to the field may show spontaneous order in a certain temperature
interval. 
\end{abstract}
\ifodd\a 

\markright{\rm \emph{Philosophical Magazine B}, submitted (1999)}

\thispagestyle{myheadings}\pagestyle{myheadings}\else

Contact author: \\
A. Hucht, Theoretische Tieftemperaturphysik, \\
Gerhard-Mercator-Universität Duisburg, D-47048 Duisburg \\
Tel: x49-203-379-3486, Fax: x49-203-379-2965, \\
Email: fred@thp.Uni-Duisburg.DE

\fi

\section{Introduction}

Experimentally it became possible in recent years to grow epitaxial thin
films of ferromagnetic materials on non-magnetic substrates with a very
high quality. This offers the possibility to stabilize crystallographic
structures which are not present in nature, and which may exhibit new properties
of high technological impact. To understand the magnetic structure of these
systems is a challenging problem both experimentally and theoretically.

One of the very interesting effects one observes in ultra-thin ferromagnetic
films is a reorientation of the spontaneous magnetization by varying either
the film thickness or the temperature. For not too thin films the magnetization
generally is in-plane due to the dipole interaction (shape anisotropy).
On the other hand in very thin films this may change due to various competing
anisotropy energies in these materials. At the surfaces of the film due
to the broken symmetry uniaxial anisotropy energies which may favor a perpendicular
magnetization may develop\cite{Neel54} or in the inner layers of the film
due to strain induced distortion bulk anisotropy energies may occur absent
or very small in the ideal crystal.

A perpendicular anisotropy has been observed for instance for ultra-thin
Fe-films grown on Ag(100) where in the ground state the magnetization is
perpendicular to the film\cite{Allen94,Pappas90,Qiu93}. Increasing the
temperature or the film thickness the magnetization stays perpendicular
until at a certain temperature it starts canting reaching finally an in-plane
orientation for temperatures well below the ordering temperature \( T_{c} \).

A different type of spin reorientation transition (SRT) occurs in Ni grown
on Cu(001). Very thin films then show a tetragonal distortion resulting
in a stress-induced uniaxial anisotropy energy in the inner layers with
its easy axis perpendicular to the film. If this is strong enough the magnetization
is in-plane in the ground state but switches to an orientation perpendicular
to the film for higher temperature or film thickness\cite{Schulz94}.

Phenomenologically in order to describe the SRT the energy (or the free
energy at finite temperatures) is expanded in terms of the orientation
of the magnetization vector relative to the film introducing temperature
dependent anisotropy coefficients \( K_{i}(T) \) compatible with the underlying
symmetry of the film. The temperature dependence of these coefficients
is then studied experimentally (for a review see\cite{Gradmann93}). Theoretically,
it has been shown by various groups that the physics of the SRT can be
understood quite well within the framework of statistical spin models\cite{Taylor93,Moschel94,Hucht95,Hucht96}.
Note that both types of SRTs observed experimentally can be explained within
the same approach\cite{Jensen96,Hucht97} and that the anisotropy coefficients
\( K_{i}(T) \) can be calculated\cite{Hucht97}.

\section{The model}

In the present paper we focus on the dependence of the SRT on external
magnetic fields. The calculations are done in the framework of a classical
ferromagnetic Heisenberg model on a simple cubic lattice consisting of
\( L \) two-dimensional layers with the \( \vec{z} \)-direction normal
to the film. The Hamiltonian reads 
\begin{eqnarray}
\mathcal{H} & = & -\frac{J}{2}\sum _{\langle ij\rangle }\vec{s}_{i}\cdot \vec{s}_{j}-\sum _{i}D_{\lambda _{i}}(s_{i}^{z})^{2}-\sum _{i}\vec{B}\cdot \vec{s}_{i}\nonumber \label{hami} \\
 & + & \frac{\omega }{2}\sum _{ij}r_{ij}^{-3}\vec{s}_{i}\cdot \vec{s}_{j}-3r_{ij}^{-5}(\vec{s}_{i}\cdot \vec{r}_{ij})(\vec{r}_{ij}\cdot \vec{s}_{j}),\label{e:H} 
\end{eqnarray}
 where \( \vec{s}_{i}=(s_{i}^{x},s_{i}^{y},s_{i}^{z}) \) are spin vectors
of unit length at position \( \vec{r}_{i}=(r_{i}^{x},r_{i}^{y},r_{i}^{z}) \)
in layer \( \lambda _{i} \) and \( \vec{r}_{ij}=\vec{r}_{i}-\vec{r}_{j} \).
\( J \) is the nearest-neighbor exchange coupling constant, \( D_{\lambda } \)
is the local uniaxial anisotropy which depend on the layer index \( \lambda =1\ldots L \),
\( \vec{B} \) denotes the external magnetic field with the effective magnetic
moment \( \mu  \) of the spins incorporated, and \( \omega =\mu _{0}\mu ^{2}/4\pi a^{3} \)
is the strength of the long range dipole interaction on a lattice with
lattice constant \( a \) (\( \mu _{0} \) is the magnetic permeability).
All energies and temperatures are measured in units of \( \omega  \),
and we set \( k_{\mathrm{B}}=1 \). Assuming translational invariance of
the Hamiltonian the local magnetization in mean field theory only depends
on the layer index, i.e. \( \langle \vec{s}_{i}\rangle =\vec{m}_{\lambda } \)
if \( \vec{s}_{i} \) is a spin in layer \( \lambda  \). A molecular field
approximation for the Hamiltonian (\ref{e:H}) results in \( L \) effective
one particle Hamiltonians from which the free energy functional can be
obtained. For more details the reader is referred to Ref.\cite{Hucht97}.

For appropriate sets of parameters a SRT in zero magnetic field from perpendicular
orientation at low temperatures to in-plane orientation at high temperatures
is found\cite{Hucht96}. In this contribution we will investigate the influence
of an applied field on this transition, which has only been addressed in
the framework of a phenomenological approach\cite{Millev98b} until now.

\section{Results}

\ifodd\a
\begin{figure}
[t]

{\par\centering \resizebox*{12cm}{!}{\includegraphics{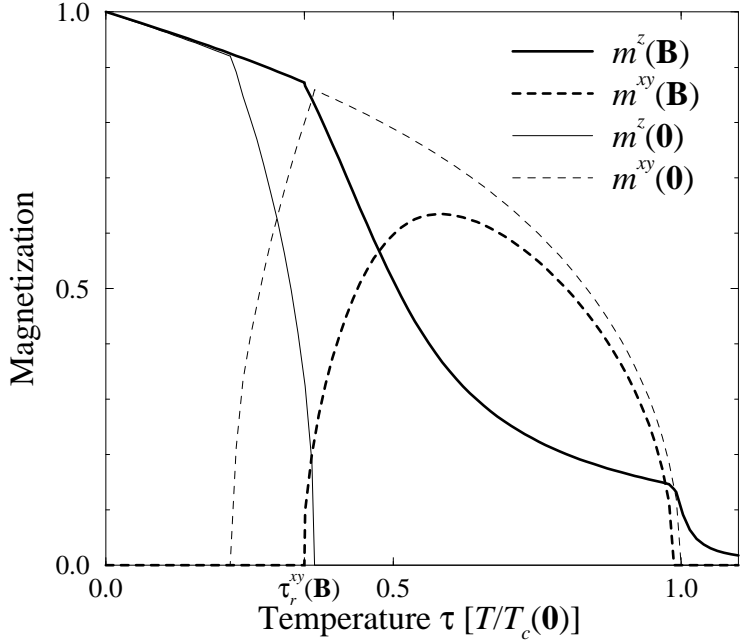}} \par}

\caption{Magnetization components \protect\( m^{xy}\protect \) and \protect\( m^{z}\protect \)
of a bilayer in a small perpendicular magnetic field \protect\( \vec{B}/\omega =0.6\, \hat{\vec{z}}\protect \)
versus reduced temperature \protect\( \tau \protect \). The thin lines
are calculated for zero magnetic field. \label{f:Mag-Bz}}
\end{figure}
\else {[}\noun{Insert Fig.~\ref{f:Mag-Bz} here}{]} \fi Figure~\ref{f:Mag-Bz}
shows the magnetization components of the total magnetization \( \vec{m}=(\vec{m}_{1}+\vec{m}_{2})/2 \)
as function of reduced temperature for an applied field \( \vec{B}/\omega =0.6\, \hat{\vec{z}} \)
(\( \hat{\vec{z}} \) is the unit vector normal to the film). A bilayer
is considered with an exchange interaction \( J/\omega =60 \) and uniaxial
anisotropies \( D_{1}/\omega =12 \) in the first layer and \( D_{2}/\omega =3 \)
in the other layer. This system has already been studied in zero magnetic
field\cite{Hucht96} and it is sufficient to describe the continuous transition.
With this set of parameters a SRT from perpendicular orientation at low
temperature to in-plane orientation at high temperatures is obtained in
zero magnetic field (thin lines in figure~\ref{f:Mag-Bz}). In a perpendicular
magnetic field \( \vec{B}\Vert \hat{\vec{z}} \) the component of the magnetization
\( m^{z} \) parallel to the field is finite for all temperatures. Nevertheless,
at a temperature \( \tau _{r}^{xy}(\vec{B})>\tau _{r}^{xy}(\vec{0}) \)
an in-plane spontaneous magnetization \( m^{xy} \) appears as in the field
free case. This in-plane phase \( \mathrm{FM}_{xy} \) is \emph{not} suppressed
by the perpendicular external magnetic field. We observe a quite interesting
reentrant phase transition: a spontaneous in-plane magnetization appears
when \emph{increasing} the temperature, and it vanishes at a higher temperature
\( \tau _{c}(\vec{B}) \) sightly below the Curie temperature of the field
free case \( \tau _{c}(\vec{0})=1 \).

\ifodd\a
\begin{figure}
[t]

{\par\centering \resizebox*{12cm}{!}{\includegraphics{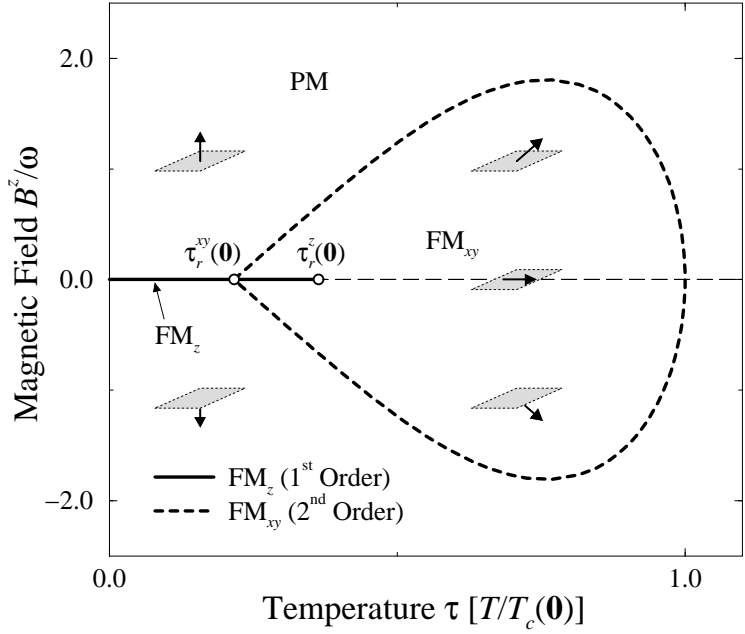}} \par}

\caption{Phase diagram in a perpendicular magnetic field \protect\( \vec{B}=B^{z}\hat{\vec{z}}\protect \).
The curves in figure~\ref{f:Mag-Bz} are calculated along the line \protect\( B^{z}/\omega =0.6\protect \).
The arrows indicate the direction of the magnetization with respect to
the film. \label{f:Phase-Bz}}
\end{figure}
\else {[}\noun{Insert Fig.~\ref{f:Phase-Bz} here}{]} \fi The corresponding
phase diagram is depicted in figure~\ref{f:Phase-Bz}. Within the area
encircled by the broken line there exists an ordered phase \( \mathrm{FM}_{xy} \)
with in-plane component of the magnetization \( m^{xy}>0 \). The sketches
of the magnetization with respect to the film show how the magnetization
rotates from \( \vec{m}=-m^{z}\hat{\vec{z}} \) inside the paramagnetic
regime \( \mathrm{PM} \) via remanence at \( \vec{m}=m^{xy}\hat{\vec{y}} \)
to \( \vec{m}=m^{z}\hat{\vec{z}} \) when the applied field is increased
from negative to positive values at a fixed temperature \( \tau >\tau _{r}^{z}(\vec{0}) \). 

The phase transition at the boundary is of second order. It is important
to note that this phase boundary continuously connects the reorientation
temperature \( \tau _{r}^{xy}(\vec{B}) \) and the Curie temperature \( \tau _{c}(\vec{B}) \)
for all perpendicular magnetic fields with \( |B^{z}|<B^{z}_{c} \) (\( B^{z}_{c}/\omega \approx 1.81 \)
for our set of parameters). As we do not find any multi-critical points
on this boundary at least in mean field theory, the critical behavior should
be the same at the whole boundary and should belong to the universality
class of the two dimensional dipolar Heisenberg model.

The ordered phase \( \mathrm{FM}_{z} \) in which the perpendicular magnetization
\( m^{z} \) shows spontaneous order is only stable at \( B^{z}=0 \) for
temperatures below the second order critical point \( \tau ^{z}_{r}(\vec{0}) \).
A first order transition is obtained when crossing this phase in a perpendicular
magnetic field. This phase is similar to the ordered phase in the two-dimensional
ferromagnetic Ising model. Due to the uniaxial anisotropy \( D_{\lambda } \)
it very likely will belong to the same universality class. 

\ifodd\a
\begin{figure}
[t]

{\par\centering \resizebox*{12cm}{!}{\includegraphics{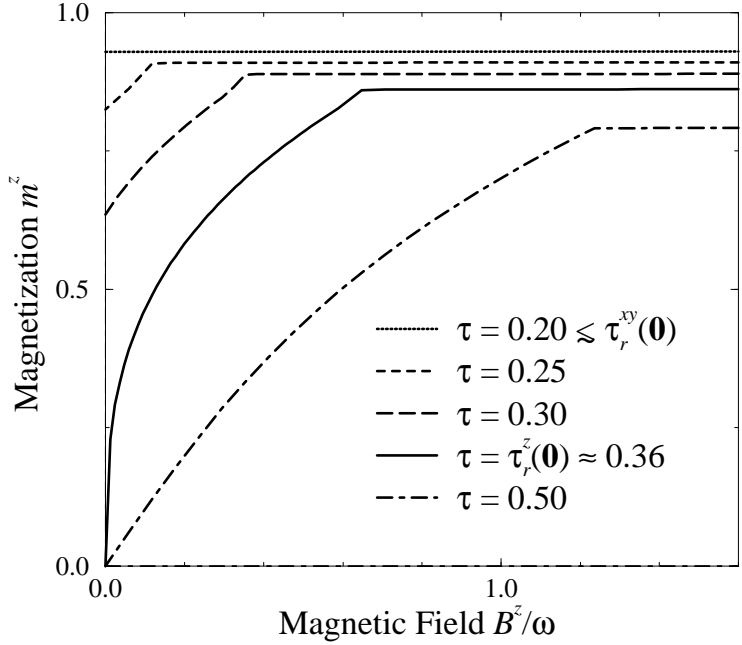}} \par}

\caption{Perpendicular magnetization \protect\( m^{z}\protect \) as function of
the perpendicular magnetic field \protect\( B^{z}/\omega \protect \) for
several reduced temperatures \protect\( \tau =T/T_{c}(\vec{0})\protect \).
\label{f:Hys}}
\end{figure}
\else {[}\noun{Insert Fig.~\ref{f:Hys} here}{]} \fi In figure~\ref{f:Hys}
we depict the field dependence of the magnetization for several temperatures
below \( \tau _{c}(\vec{0}) \). Note that this diagram is symmetric with
respect to the origin. For \( \tau <\tau _{r}^{xy}(\vec{0}) \) we find
a jump in \( m^{z} \) at zero field as we cross the phase \( \mathrm{FM}_{z} \).
This jump is still present for \( \tau _{r}^{xy}(\vec{0})<\tau <\tau _{r}^{z}(\vec{0}) \),
but now the zero-field state is canted, and with increasing field we get
a transition into the paramagnetic phase where \( \vec{m}\Vert \vec{B} \).
At \( \tau _{r}^{z}(\vec{0}) \) the \( zz \)-component of the static
zero-field susceptibility tensor \( \chi _{0}^{\mu \nu }=dm^{\mu }/dB^{\nu }|_{\vec{B}=\vec{0}} \)
diverges (note that the same holds at \( \tau _{r}^{xy}(\vec{0}) \) for
the in-plane components \( \chi _{0}^{xx} \) and \( \chi _{0}^{yy} \)),.
For \( \tau >\tau _{r}^{z}(\vec{0}) \) the perpendicular remanence is
zero, nevertheless the magnetization curve has a kink when we leave the
ordered phase \( \mathrm{FM}_{xy} \). Finally, for \( \tau >\tau _{c}(\vec{0}) \)
we find the usual paramagnetic behavior (not displayed).

\ifodd\a
\begin{figure}
[t]

{\par\centering \resizebox*{12cm}{!}{\includegraphics{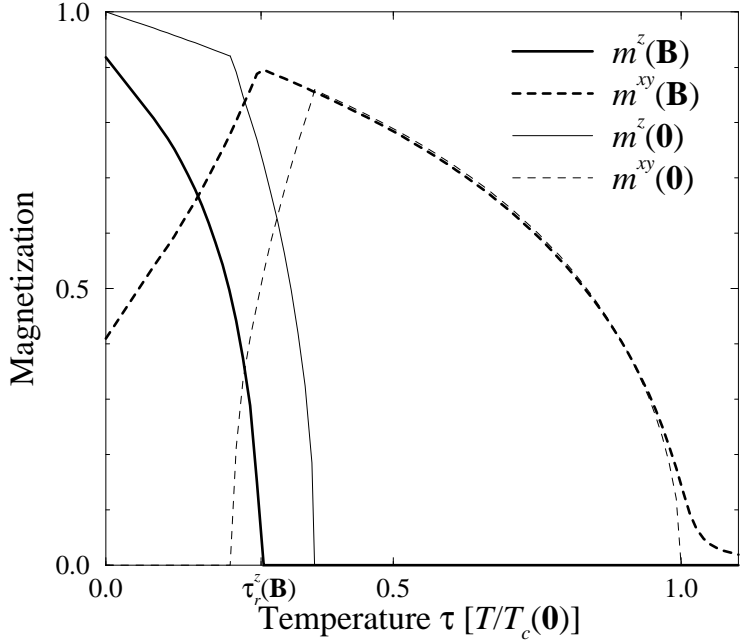}} \par}

\caption{Same scenario as in figure~\ref{f:Mag-Bz}, but with the magnetic field
\protect\( \vec{B}/\omega =0.6\, \hat{\vec{y}}\protect \) parallel to
the film. \label{f:Mag-By}}
\end{figure}
\else {[}\noun{Insert Fig.~\ref{f:Mag-By} here}{]} \fi A different scenario
is obtained in a parallel magnetic field \( \vec{B}=B^{y}\, \hat{\vec{y}} \).
Figure~\ref{f:Mag-By} shows the corresponding magnetization components.
Due to the in-plane field the in-plane ordered phase \( \mathrm{FM}_{xy} \)
is completely suppressed, the critical points \( \tau _{r}^{xy}(\vec{B}) \)
and \( \tau _{c}(\vec{B}) \) are not defined anymore. Nevertheless, the
\( z \)-component of the magnetization \( m^{z} \) shows spontaneous
order at temperatures below \( \tau _{r}^{z}(\vec{B})<\tau _{r}^{z}(\vec{0}) \),
where it vanishes continuously with a second order phase transition. Above
this temperature no spontaneous order is present in the system.

\ifodd\a
\begin{figure}
[t]

{\par\centering \resizebox*{12cm}{!}{\includegraphics{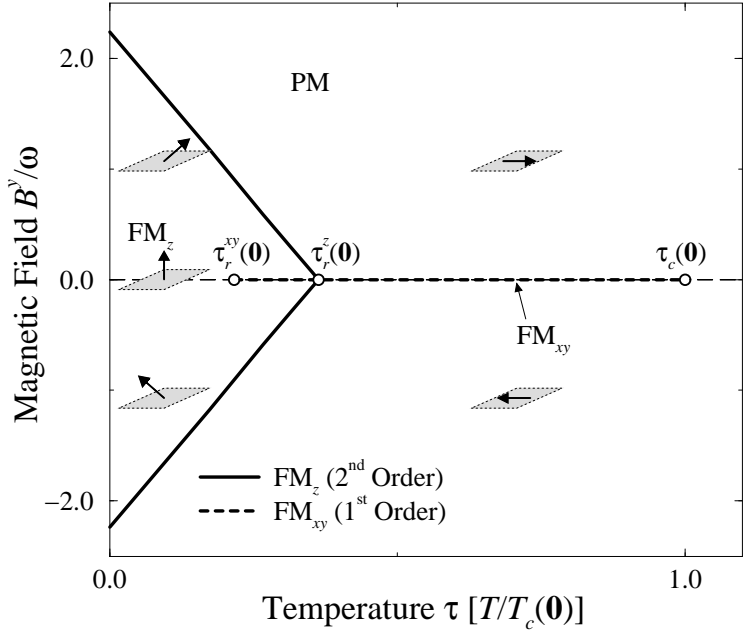}} \par}

\caption{Phase diagram in a parallel magnetic field \protect\( \vec{B}=B^{y}\hat{\vec{y}}\protect \).
The curves in figure~\ref{f:Mag-By} are calculated along the line \protect\( B^{y}/\omega =0.6\protect \).
The arrows indicate the direction of the magnetization with respect to
the film. \label{f:Phase-By}}
\end{figure}
\else {[}\noun{Insert Fig.~\ref{f:Phase-By} here}{]} \fi The corresponding
phase diagram for this situation is shown in figure~\ref{f:Phase-By}.
To the left of the solid line we find an ordered phase \( \mathrm{FM}_{z} \)
with order parameter \( m^{z}>0 \). The phase boundary is of second order
and ends at \( \tau =0 \) and \( \vec{B}=\pm B^{y}_{c}\hat{\vec{y}} \)
with \( B^{y}_{c}/\omega \approx 2.24 \) for our set of parameters. The
in-plane phase \( \mathrm{FM}_{xy} \) is only stable at \( B^{y}=0 \),
it starts at \( \tau ^{xy}_{r}(\vec{0}) \) and ends at \( \tau _{c}(\vec{0}) \)
in analogy to figure~\ref{f:Phase-Bz}. Again we sketched the orientation
of the magnetization relative to the film. 

\ifodd\a
\begin{figure}
[t]

{\par\centering \includegraphics{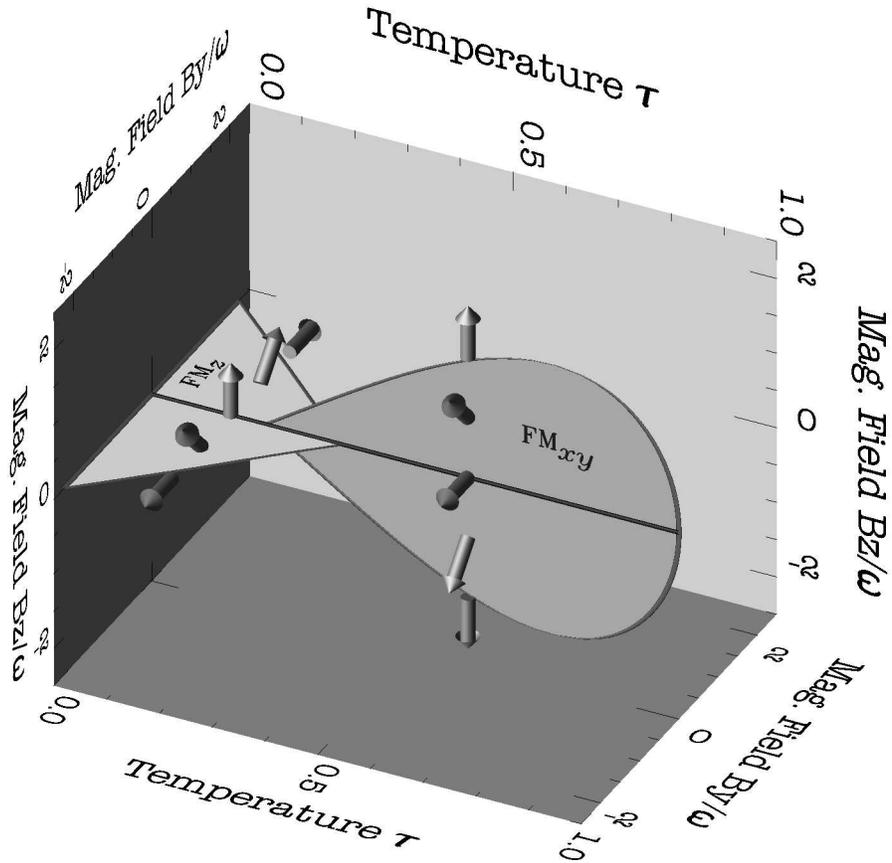} \par}

\caption{Three dimensional phase diagram in the \protect\( (\tau ,B^{y}/\omega ,B^{z}/\omega )\protect \)
parameter space. The arrows sketch the spontaneous magnetization inside
the phases \protect\( \mathrm{FM}_{z}\protect \) and \protect\( \mathrm{FM}_{xy}\protect \),
outside these phases we have \protect\( \vec{m}\Vert \vec{B}\protect \).
\label{f:Phase-3D}}
\end{figure}
\else {[}\noun{Insert Fig.~\ref{f:Phase-3D} here}{]} \fi In general
we find that the behavior with in-plane magnetic field is similar to the
case with perpendicular magnetic field if one exchanges both in-plane and
perpendicular directions and reverses the temperature. To make this symmetry
even clearer, in figure~\ref{f:Phase-3D} we depict a three dimensional
phase diagram of the system. The two phases with spontaneous magnetization
intersect along the line \( \vec{B}=\vec{0},\tau _{r}^{xy}(\vec{0})<\tau <\tau _{r}^{z}(\vec{0}) \).
In this range we find spontaneous canted magnetization.

As shown in Ref.~\cite{Hucht96}, a bilayer and thicker films may show
both second order and first order SRTs at zero field depending on the distribution
of uniaxial anisotropies \( D_{\lambda } \): If in the bilayer the deviation
\( \Delta =|D_{1}-D_{2}| \) approaches the critical value \( \Delta _{c}\approx \omega  \),
the critical points \( \tau _{r}^{xy}(\vec{0}) \) and \( \tau _{r}^{z}(\vec{0}) \)
merge into one multi-critical point where the order of the SRT in zero
field changes from second to first order. If we apply these former results
to this work, we conclude that the intersection width of the phases \( \mathrm{FM}_{z} \)
and \( \mathrm{FM}_{xy} \) is finite for \( \Delta >\Delta _{c} \) and
zero for \( \Delta \leq \Delta _{c} \).

\section{Summary}

For the Heisenberg model with dipole interaction and uniaxial anisotropy
in an external magnetic field, we calculated magnetization curves as function
of both temperature and magnetic fields in the mean field approximation
and determined phase diagrams in the temperature-field parameter space.
We find that the spin reorientation transition present in this model is
not suppressed completely by an external magnetic field, as the magnetization
component perpendicular to the field may show spontaneous order. The phase
diagrams reveil that the field dependent transition point \( T_{r}^{xy}(\vec{B}) \),
where a canted magnetization occurs with increasing temperature, and the
field dependent Curie temperature \( T_{c}(\vec{B}) \) are connected by
a continuous phase boundary. Hence both transitions should be in the same
universality class.

\subsubsection*{Acknowledgment}

This work was supported by the Deutsche Forschungsgemeinschaft through
Sonderforschungsbereich 166.

\ifodd\a\else

\newpage
\begin{figure}
[t]

\section*{Figure captions}

\bigskip{}
\caption{Magnetization components \protect\( m^{xy}\protect \) and \protect\( m^{z}\protect \)
of a bilayer in a small perpendicular magnetic field \protect\( \vec{B}/\omega =0.6\, \hat{\vec{z}}\protect \)
versus reduced temperature \protect\( \tau \protect \). The thin lines
are calculated for zero magnetic field. \label{f:Mag-Bz}}

\bigskip{}
\caption{Phase diagram in a perpendicular magnetic field \protect\( \vec{B}=B^{z}\hat{\vec{z}}\protect \).
The curves in figure~\ref{f:Mag-Bz} are calculated along the line \protect\( B^{z}/\omega =0.6\protect \).
The arrows indicate the direction of the magnetization with respect to
the film. \label{f:Phase-Bz}}

\bigskip{}
\caption{Perpendicular magnetization \protect\( m^{z}\protect \) as function of
the perpendicular magnetic field \protect\( B^{z}/\omega \protect \) for
several reduced temperatures \protect\( \tau =T/T_{c}(\vec{0})\protect \).
\label{f:Hys}}

\bigskip{}
\caption{Same scenario as in figure~\ref{f:Mag-Bz}, but with the magnetic field
\protect\( \vec{B}/\omega =0.6\, \hat{\vec{y}}\protect \) parallel to
the film. \label{f:Mag-By}}

\bigskip{}
\caption{Phase diagram in a parallel magnetic field \protect\( \vec{B}=B^{y}\hat{\vec{y}}\protect \).
The curves in figure~\ref{f:Mag-By} are calculated along the line \protect\( B^{y}/\omega =0.6\protect \).
The arrows indicate the direction of the magnetization with respect to
the film. \label{f:Phase-By}}

\bigskip{}
\caption{Three dimensional phase diagram in the \protect\( (\tau ,B^{y}/\omega ,B^{z}/\omega )\protect \)
parameter space. The arrows sketch the spontaneous magnetization inside
the phases \protect\( \mathrm{FM}_{z}\protect \) and \protect\( \mathrm{FM}_{xy}\protect \),
outside these phases we have \protect\( \vec{m}\Vert \vec{B}\protect \).
\label{f:Phase-3D}}
\end{figure}

~

\clearpage

\newpage\vspace*{\fill}
{\par\centering \resizebox*{1\textwidth}{!}{\includegraphics{Mag-Bz=0.01-en.eps}} \par}
\vfill{}

Figure \ref{f:Mag-Bz}

\newpage\vspace*{\fill}
{\par\centering \resizebox*{1\textwidth}{!}{\includegraphics{Phase-Bz-T-L=2-en.eps}} \par}
\vfill{}

Figure \ref{f:Phase-Bz}

\newpage\vspace*{\fill}
{\par\centering \resizebox*{1\textwidth}{!}{\includegraphics{Mag-Bz-en-nometa.eps}} \par}
\vfill{}

Figure \ref{f:Hys}

\newpage\vspace*{\fill}
{\par\centering \resizebox*{1\textwidth}{!}{\includegraphics{Mag-By=0.01-en.eps}} \par}
\vfill{}

Figure \ref{f:Mag-By}

\newpage\vspace*{\fill}
{\par\centering \resizebox*{1\textwidth}{!}{\includegraphics{Phase-By-T-L=2-en.eps}} \par}
\vfill{}

Figure \ref{f:Phase-By}

\newpage\vspace*{\fill}
{\par\centering \resizebox*{1\textwidth}{!}{\includegraphics{phase-3D-bw-en.eps}} \par}
\vfill{}

Figure \ref{f:Phase-3D}

\fi

\end{document}